\documentclass[structabstract]{aa}

\usepackage{graphicx}
\usepackage{rotating}
\usepackage{txfonts}

\def\dsct{$\delta$~Scuti }

\def\Msun{$M_{\odot}$}
\def\Lsun{$L_{\odot}$}
\def\Rsun{$R_{\odot}$}

\def\Teff{\ensuremath{T_{\mathrm{eff}}}}
\def\cd{d$^{\rm -1}$}
\def\logg{\ensuremath{\log g}}
\def\vmic{$\upsilon_{\mathrm{mic}}$}

\def\vsini{\ensuremath{{\upsilon}\sin i}}
\def\kms{$\mathrm{km\,s}^{-1}$}

\def\llm{{\sc LLmodels}}

\def\vald{{\sc VALD}}

\begin{document}

\title{Refining the asteroseismic model for the young $\delta$ Scuti star HD\,144277 using HARPS spectroscopy\thanks{This work is based on ground-based observations made with the 3.6m telescope
at La Silla Observatory under the ESO Large Programme LP185.D-0056}}

\author{K. Zwintz\inst{1} \and
T. Ryabchikova\inst{2} \and
P. Lenz\inst{3} \and
A. A. Pamyatnykh\inst{2,3,4} \and
L. Fossati\inst{5} \and
T. Sitnova\inst{2,6} \and
M. Breger\inst{7} \and
E. Poretti\inst{8} \and
M. Rainer\inst{8} \and
M. Hareter\inst{4} \and
L. Mantegazza\inst{8}
}

\offprints{K. Zwintz, \\ \email{konstanze.zwintz@ster.kuleuven.be}}

\institute{
       Instituut voor Sterrenkunde, K. U. Leuven, Celestijnenlaan 200D, B-3001 Leuven, Belgium\\
       \email: konstanze.zwintz@ster.kuleuven.be \and
   Institute of Astronomy, Russian Academy of Sciences, Pyatnitskaya Str 48, 109017 Moscow, Russia \and
Copernicus Astronomical Centre, Bartycka 18, 00-716 Warsaw, Poland \and
   University of Vienna, Institute of Astronomy, T\"urkenschanzstrasse 17, A-1180 Vienna, Austria \and
   Argelander-Institut f\"ur Astronomie der Universit\"at Bonn, Auf dem H\"ugel 71, 53121 Bonn, Germany \and
Moscow M.V. Lomonosov State University, Sternberg Astronomical Institute, Universitetskii pr. 13, Moscow, 119992 Russia \and
   Dept. of Astronomy, University of Texas at Austin, Austin, TX 78712, USA \and
   INAF-Osservatorio Astronomico di Brera, Merate, Italy
}

\date{Received / Accepted }

\abstract
{HD\,144277 was previously discovered by Microvariability and Oscillations of Stars (MOST) space photometry to be a young and hot \dsct star showing regular groups of pulsation frequencies. The first asteroseismic models required lower than solar metallicity to fit the observed frequency range based on a purely photometric analysis.}
{The aim of the present paper is to determine, by means of high-resolution spectroscopy, fundamental stellar parameters required for the asteroseismic model of HD\,144277, and subsequently, to refine it. }
{High-resolution, high signal-to-noise spectroscopic data obtained with the HARPS spectrograph were used to determine the fundamental parameters and chemical abundances of HD\,144277. These values were put into context alongside the results from asteroseismic models.}
{The effective temperature, \Teff, of HD\,144277 was determined as 8640\,$^{+300}_{-100}$\,K, \logg\ is 4.14 $\pm$ 0.15 and the projected rotational velocity, \vsini, is 62.0 $\pm$ 2.0 \kms. As the \vsini\ value is significantly larger than previously assumed, we refined the first asteroseimic model accordingly. The overall metallicity $Z$ was determined to be 0.011 where the light elements He, C, O, Na, and S show solar chemical composition, but the heavier elements are significantly underabundant. In addition, the radius of HD\,144277 was determined to be 1.55 $\pm$ 0.65\Rsun\ from spectral energy distribution (SED) fitting, based on photometric data taken from the literature.}
{From the spectroscopic observations, we could confirm our previous assumption from asteroseismic models that HD\,144277 has less than solar metallicity. The fundamental parameters derived from asteroseismology, \Teff, \logg, L/\Lsun\ and R/\Rsun\ agree within one sigma to the values found from spectroscopic analysis. As the \vsini\ value is significantly higher than assumed in the first analysis, near-degeneracies and rotational mode coupling were taken into account in the new models. These suggest that HD\,144277 has an equatorial rotational velocity of about 80\kms\ and is seen equator-on. The observed frequencies are identified as prograde modes. }

\keywords{stars: variables: $\delta$ Sct - stars: oscillations - stars: individual: HD\,144277 - techniques: photometric}

\maketitle
\titlerunning{Refined asteroseismic model of HD\,144277}
\authorrunning{K. Zwintz et al.}

\section{Introduction}

High-precision time-series photometry obtained by the Microvariability and Oscillations of Stars (MOST; Walker et al. \cite{wal03}) space telescope during two consecutive years discovered \dsct\ pulsations in HD\,144277 (Zwintz et al. \cite{zwi11}; Paper I, hereafter). The twelve independent pulsation frequencies identified lie in four distinct groups, i.e., they show regular frequency patterns. The first asteroseismic analysis was based on limited additional knowledge of the physical parameters of HD\,144277. 
The star was only known to have a spectral type of A1V with a respective effective temperature of 9230~K (both taken from the Tycho-2 Spectral Type Catalog; Wright et al. \cite{wri03}), and a parallax of 7.00$\pm$5.59 mas (Kharchenko \& Roeser \cite{kha09}).
No spectroscopic data were available and the location of HD\,144277 at the hot border of the \dsct\ instability strip was determined by dedicated Str\"omgren photometry (see Paper I for more details). 

The asteroseismic analysis conducted in Paper I found evidence of two radial modes of sixth and seventh overtones under the assumption of slow rotation. The models calculated in Paper I needed slightly less than solar metallicity and a moderate enhancement of the helium abundance compared to the standard solar chemical composition. 

With the lack of high-resolution spectroscopy for HD\,144277, it was not possible to test if the non-solar abundances required from asteroseismology for the excitation of the observed \dsct\ frequencies could also be found in the atmosphere of the star. The asteroseismic models would of course also be different, if the \vsini\ were significantly larger than assumed in Paper I. We obtained high-resolution spectra for HD\,144277 in order to test the validity of the first asteroseismic models and the assumptions needed to explain the observed frequencies.

We present here the results of our spectral analysis for HD~144277 that lead to a refinement of the asteroseismic models first presented in Paper I.

\section{High-resolution spectroscopy: observations and data reduction}

Fundamental parameters and abundances for HD\,144277 were determined from spectra obtained during five nights between 2011 June 28 and July 3 with the HARPS spectrograph (Mayor et al. \cite{may03}) at the 3.6\,m telescope of ESO La Silla Observatory. 
HD 144277 was used as a backup target for the ESO LP185.D-0056 to extend the physical scenario of the $\delta$ Sct stars observed with CoRoT.
In the adopted EGGS configuration the fiber-fed high-resolution \'echelle spectrograph has a resolving power of 80\,000.
Each spectrum covers the wavelength range of 3781 -- 6913\,\AA.

\begin{figure}[htb]
\centering
\includegraphics[width=0.50\textwidth,clip]{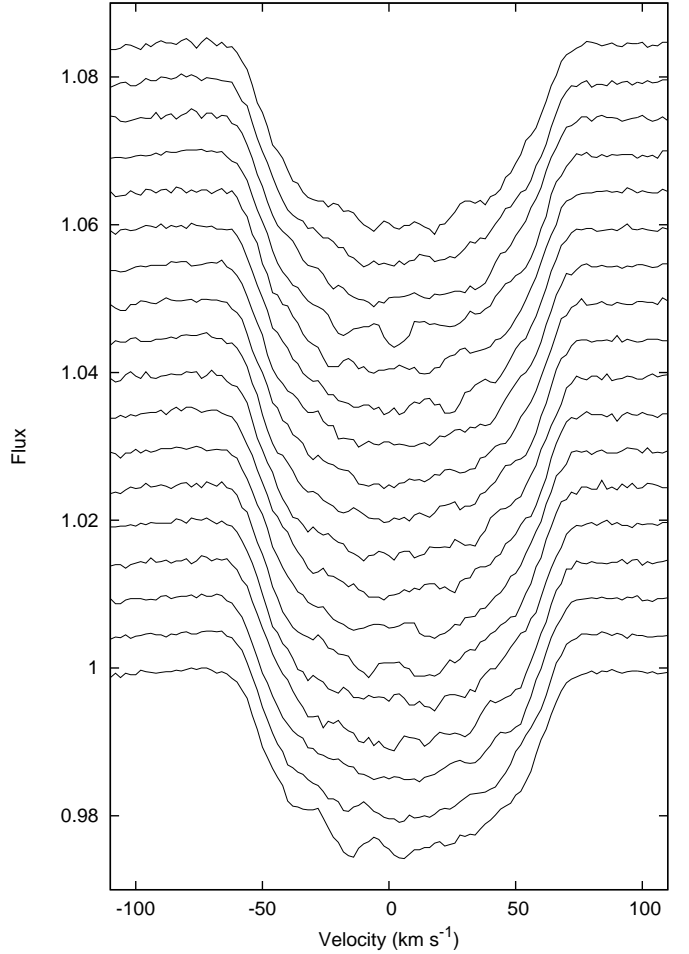}
\caption{Mean line profiles for each of the 18 single spectra obtained for HD\,144277 on different nights. Spectral intensities (in continuum units) on the y-axis are arbitrarily shifted for better visibility. Spectra are in chronological order from bottom to top.  }
\label{lpv}
\end{figure}

The 18 single spectra of HD\,144277 were observed for exposure times of 1100 (15 spectra), 1200 (two spectra) and 1300 (one spectrum) seconds and have mean pixel-by-pixel signal-to-noise ratios (S/N) between 143 and 248 calculated in the wavelength range of 5805 to 5825\,\AA. The HARPS spectra, in the adopted configuration, cover the hydrogen Balmer lines H$\alpha$, H$\beta$, H$\gamma$, H$\delta$, and H$\epsilon$.
Since the 18 single spectra were obtained during five different nights and since the pulsation periods discovered earlier are shorter than 24 minutes, the data could not be used as a spectroscopic time series to study line profile variations. 
Figure \ref{lpv} shows that both the radial velocity variations and the line profile variations are much smaller than the rotational broadening. 
The 18 single spectra were therefore combined to increase the S/N of the combined spectrum that then amounts to 363 computed again in the wavelength range of 5805 to 5825\,\AA.

The spectra were reduced using both the ESO pipeline and a semi-automatic MIDAS pipeline (Rainer \cite{rai03}). They were then normalized by fitting a low-order polynomial to carefully selected continuum points. 
The resulting normalized combined high-resolution spectrum was thereafter used to determine the fundamental stellar parameters presented in this article.

\section{Atmospheric parameters determination and abundance analysis}

The starting values for determining the fundamental parameters and chemical abundances of HD\,144277 were taken from multi-color photometry.
For HD\,144277 Str\"omgren colors (Paper I) without $\beta$-parameter, and usual broad-band photometric data are available. With three different calibrations of Str\"omgren photometric indices (Moon \& Dworetsky \cite{MD}, Balona \cite{Balona94}, Ribas et al. \cite{Ribas97}) included in the {\sc TEMPLOGG} package (Kaiser \cite{TMG}), and with a proper range of assumed $\beta$-parameters, we estimated \Teff=8800$\pm$150\ K, \logg=3.9 -- 4.1, and metallicity $M$ of -0.28 from photometry.

A comparison between the observed spectrum of HD\,144277 and the synthetic spectrum, calculated for an atmospheric model with the above mentioned parameters, shows a rather high rotational velocity, and, as a consequence, strong line-blending. 
This led us to use spectral synthesis and employ the SME (Spectroscopy Made Easy) package to derive the atmospheric parameters for HD\,144277. The SME software was developed by Valenti \& Piskunov (\cite{sme}) and, for example, was successfully applied to determine the atmospheric parameters of FGK stars (Valenti \& Fischer \cite{VF}). SME allows to derive effective temperature, surface gravity, overall metallicity, individual element abundances, and the microturbulent, macroturbulent, rotational, and radial velocities of a star by fitting synthetic spectra to the observed ones. Spectral synthesis calculations may be performed for different grids of model atmospheres and are interpolated between the grid nodes. 
We used a model grid calculated with the \llm\ stellar model atmosphere code (Shulyak et al. \cite{llm}) for microturbulent velocity \vmic=2.0~\kms, which ranges from 4500--22000~K in effective temperature, from 2.5--5.0~dex in surface gravity, and from -0.8 -- 0.8~dex in metallicity (see Table 5 in Tkachenko et al. \cite{LL-grid}). The corresponding steps are 0.1 in \logg, 0.1~dex in metallicity, 100~K in the effective temperature region from 4500--10000~K and 250~K for higher \Teff\ values.

The following spectral regions were chosen for the fitting procedure of HD\,144277: 4140-4410~\AA, 4411-4705~\AA, 4750-4963~\AA, 4963-5298~\AA, 5257-5620~\AA, 6100-6250~\AA, 6340-6465~\AA\, and 6480-6580~\AA. These ranges include the three Balmer lines H$\alpha$, H$\beta$, and H$\gamma$. Atomic parameters were extracted from the Vienna Atomic Line Database (\vald). 
We chose three different steps to derive the star's parameters: (i) we used only the three Balmer lines as input for SME, (ii) we took the spectral regions without the Balmer lines, and (iii) all above-mentioned spectral regions were selected at the same time. The corresponding solutions are  8558\,K / 4.15 / -0.56 (i), 9035\,K / 4.29 / -0.33 (ii) and 8641\,K / 4.14 / -0.54 (iii) for \Teff, \logg, and metallicity [Fe/H]. All three solutions give \vsini\ values close to 62~\kms. 
Microturbulent velocity varied around 2.5~\kms.
In the temperature region of the assumed position of HD\,144277, the hydrogen lines are rather insensitive to variations in \Teff\, and more sensitive to variations in \logg, while the ionization balance for \ion{Fe}{i}/\ion{Fe}{ii} is sensitive both to \Teff\, and \logg. The use of the ionization balance requires accurate transition probabilities for \ion{Fe}{i} and \ion{Fe}{ii} lines as well as non-local thermodynamic equilibrium (NLTE) line formation. 

\begin{figure}[htb]
\centering
\includegraphics[width=0.45\textwidth,clip]{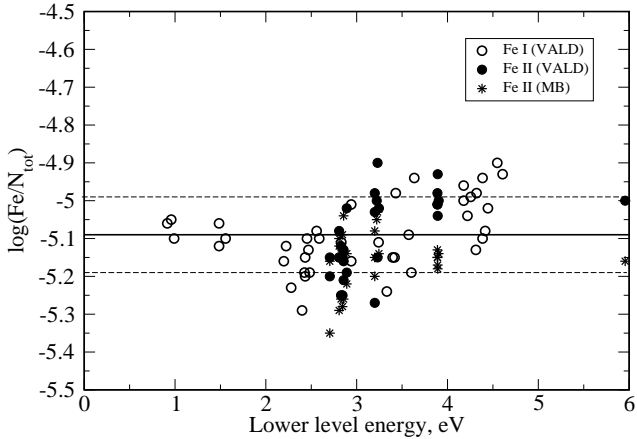}
\caption{Iron abundance based on NLTE calculations versus lower level energy. Individual abundances from \ion{Fe}{i} (open circles), \ion{Fe}{ii} (\vald\, oscillator strengths -- filled circles) and \ion{Fe}{ii} (Mel\'endez \& Barbuy \cite{MB} oscillator strengths -- asterisks) are shown. Horizontal lines indicate the mean Fe abundance with $\pm$0.1 deviation.}
\label{Fe-lines}
\end{figure}

\begin{figure}[htb]
\centering
\includegraphics[width=0.50\textwidth,clip]{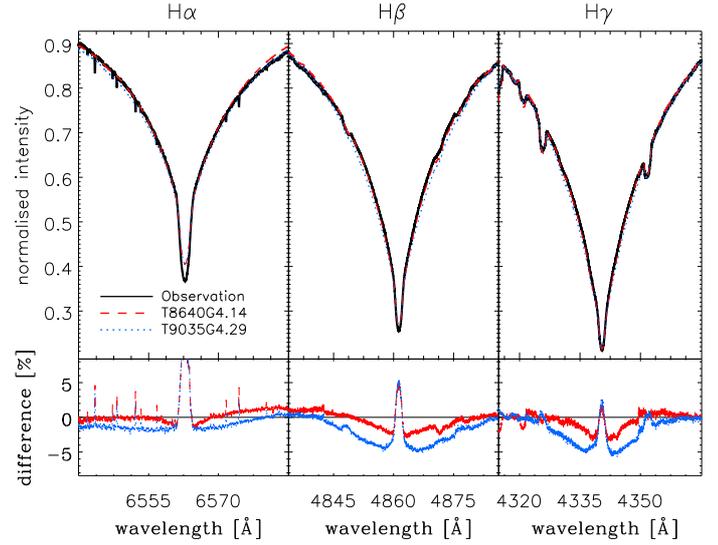}
\caption{Region of the H$\alpha$ (left), H$\beta$ (middle), and H$\gamma$ (right) lines for HD\,144277: observed 
spectra (black solid line), synthetic spectra with the final adopted stellar
parameters (\Teff=8640\,K, \logg=4.14; red solid line) and synthetic spectra using the \Teff\ and \logg\ values derived from fitting the spectral regions excluding the Balmer lines (i.e., \Teff=9035\,K, \logg=4.29; blue dashed lines). The lower panels show the differences (in \%) between observed and synthetic spectra for the three Balmer line regions.}
\label{h-lines}
\end{figure}

\begin{figure}[htb]
\centering
\includegraphics[width=0.5\textwidth,clip,angle=180]{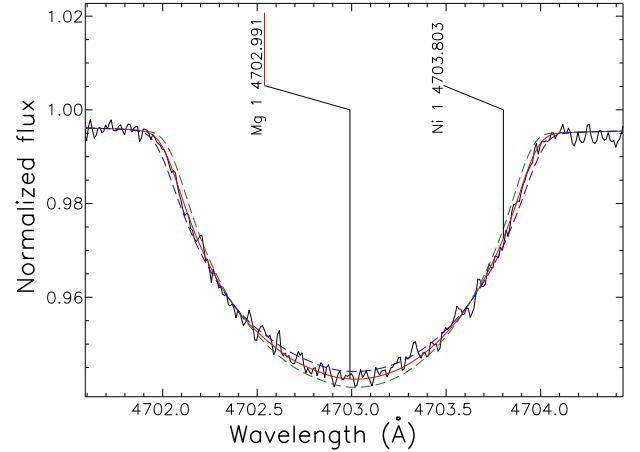}
\caption{Zoom into the region of the Mg I line at 4702\,\AA: observed spectrum (black solid line), synthetic spectrum with the best fitting \vsini\ value of 62\,\kms (red solid line) and two synthetic spectra calculated by increasing and decreasing \vsini\ by 2\kms\ (dashed blue and green lines).}
\label{vsinifit}
\end{figure}

We applied an extensive model atom of \ion{Fe}{i} and \ion{Fe}{ii} from Mashonkina (\cite{Mashonkina11}). NLTE level populations were calculated using a revised version of the DETAIL code developed by Butler and Giddings (1985). NLTE leads to a weakening of the \ion{Fe}{i} lines, and the NLTE abundance corrections vary from 0.00 to +0.09 dex. For the \ion{Fe}{ii} lines, the NLTE abundance corrections are negative and do not exceed 0.01 dex in absolute value.
The iron abundance computed from the \ion{Fe}{ii} lines depends on the adopted system of the oscillator strengths. Fig.~\ref{Fe-lines} shows the abundances derived from NLTE calculations of the \ion{Fe}{i} and \ion{Fe}{ii} lines with different oscillator strengths taken from \vald\, and from Mel\'endez \& Barbuy (\cite{MB}). While both sets provide reasonable agreement for the lines with the excitation potential $<$\,3.5~eV, a difference $\sim$0.1~dex appears for the lines with the excitation potential $>$\,3.8~eV. As a reasonable compromise we therefore adopted the final parameters for HD\,144277 as $\Teff=8640^{+300}_{-100}$\,K, \logg=4.14$\pm$0.15, \vmic=2.5$\pm$0.3~\kms, and \vsini=62$\pm$2~\kms. 
Errors for $\Teff$ are computed with SME using different spectral regions and for $\logg$ by allowing $\pm$ 0.1~dex difference in the \ion{Fe}{i}/\ion{Fe}{ii} ionization balance. 
A comparison between the observed and calculated line profiles for H$\alpha$, H$\beta$, and H$\gamma$ is shown in Fig.~\ref{h-lines}. A zoom into the Mg I line at 4702\AA\, (Fig.~\ref{vsinifit}) illustrates the accuracy of the \vsini\ determination.

Table~\ref{abun} lists the element abundances in the atmosphere of HD\,144277 with their standard deviations. The most realistic estimate of the error comes from the Fe lines because it is the only element with a statistically significant number of lines. 
For a few elements other than Fe, abundances were also derived taking NLTE effects into account. For example, the oxygen abundance was calculated following the procedure described by Sitnova et al. (\cite{SM2013}).

Our model provides us with an ionization equilibrium for Fe (see above), Mg and Ca. The Mg abundance from \ion{Mg}{i} lines was derived using NLTE calculations by Mashonkina (\cite{Mg-NLTE}). The Mg abundance in LTE computed from the \ion{Mg}{ii} lines was derived using the two lines at 4390.5 and 4481.2~\AA.

The NLTE formation of \ion{Mg}{ii} lines in the atmospheres of three A-type stars was studied by Przybilla
et al. (\cite{Mg2-NLTE}); two models were computed for supergiants, and the third model is for a main sequence star of \Teff = 9550\,K and \logg\,=\,3.95. The effective temperature of this 1D model is hotter by 900~K than that of HD~144277, but model atmospheres for both stars have comparable gravity and metallicity. Hence, we can apply these theoretical calculations to estimate possible NLTE corrections for HD~144277. 

In the model atmosphere of a star with \Teff = 9550\,K and \logg\,=\,3.95, the line at 4390~\AA\, is not affected by NLTE while the NLTE correction is -0.2~dex for the 4481.2~\AA\ line (Przybilla et al. \cite{Mg2-NLTE}).  The Mg abundance inferred from the 4390~\AA\, line is $\log (Mg/N_{\rm tot})$=-4.88.  Applying this correction to $\log (Mg/N_{\rm tot})$=-4.75 derived from the 4481.2~\AA\ line, we estimate an average Mg abundance of -4.92$\pm$0.05 from NLTE calculations that agrees well with the Mg abundance derived from the NLTE analysis of the \ion{Mg}{i} lines. 

The \ion{Ca}{i}, \ion{Ca}{ii} abundances were calculated for the atmosphere of HD~144277 in NLTE with the model atom from Mashonkina, Korn \& Przybilla (\cite{mash_ca}). 
For the \ion{Cr}{ii} lines we used the most recent transition probability calculations by Kurucz\footnote{http://kurucz.harvard.edu/atoms/2401/} that provide reasonable agreement with the abundances derived from \ion{Cr}{i} lines.

\begin{table}[ht]
\caption[ ]{LTE atmospheric abundances in HD~144277 with the error
estimates based on the internal scattering from the number of analyzed lines,
$n$.}
\label{abun}
\begin{center}
\begin{tabular}{l|lrr|c}
\hline
\hline
Ion &\multicolumn{3}{|c|}{HD~144277} &   Sun \\
   &$\log (N_{el}/N_{\rm tot})$ & $n$ &[$(N_{el}$] &$\log (N_{el}/N_{\rm tot})$  \\       \hline
\ion{He}{i }  & ~~$-$1.05:         &  1 & 0.06 & ~~$-$1.11~ \\
\ion{C}{i }   & ~~$-$3.68$\pm$0.08 &  8 &-0.07 & ~~$-$3.61~ \\
\ion{O}{i }*  & ~~$-$3.34$\pm$0.06 &  3 & 0.01 & ~~$-$3.35~ \\
\ion{Mg}{i}*  & ~~$-$4.94$\pm$0.11 &  5 &-0.50 & ~~$-$4.44~ \\
\ion{Mg}{ii}* & ~~$-$4.92$\pm$0.05 &  2 &-0.48 & ~~$-$4.44~ \\
\ion{Si}{ii}  & ~~$-$5.02$\pm$0.03 &  4 &-0.49 & ~~$-$4.53~ \\
\ion{S}{i}    & ~~$-$4.99$\pm$0.14 &  5 &-0.07 & ~~$-$4.92~ \\
\ion{Ca}{i}*  & ~~$-$6.23$\pm$0.06 & 10 &-0.53 & ~~$-$5.70~ \\
\ion{Ca}{ii}* & ~~$-$6.24$\pm$0.07 &  3 &-0.54 & ~~$-$5.70~ \\
\ion{Sc}{ii}  & ~~$-$9.49$\pm$0.04 &  3 &-0.60 & ~~$-$8.89~ \\
\ion{Ti}{ii}  & ~~$-$7.62$\pm$0.12 & 13 &-0.53 & ~~$-$7.09~ \\
\ion{Cr}{i}   & ~~$-$7.05$\pm$0.09 &  5 &-0.65 & ~~$-$6.40~ \\
\ion{Cr}{ii}  & ~~$-$6.91$\pm$0.06 & 10 &-0.51 & ~~$-$6.40~ \\
\ion{Fe}{i}*  & ~~$-$5.09$\pm$0.09 & 40 &-0.55 & ~~$-$4.54~ \\
\ion{Fe}{ii}* & ~~$-$5.09$\pm$0.10 & 29 &-0.55 & ~~$-$4.54~ \\
\ion{Ni}{i}   & ~~$-$6.38$\pm$0.10 &  3 &-0.56 & ~~$-$5.82~ \\
\ion{Sr}{ii}  & ~~$-$9.82$\pm$0.08 &  2 &-0.65 & ~~$-$9.17~ \\
\ion{Y}{ii}   & ~$-$10.43$\pm$0.10 &  2 &-0.60 & ~~$-$9.83~ \\
\ion{Ba}{ii}  & ~$-$10.53$\pm$0.10 &  4 &-0.67 & ~~$-$9.86~ \\
\hline %
\hline
\end{tabular}
\end{center}
\tablefoot{Elements treated in NLTE are marked by asterisks. [$(N_{el}/N_{\rm tot})$] is the abundance of a certain element relative to the total abundance of all elements, which is set to be equal to 1. For purpose of comparison, the last column gives the abundances of the solar atmosphere
calculated by Asplund et al. (\cite{asplund09}).}
\end{table}

HD\,144277 has solar chemical composition for the light elements C, O, Na, and S, but shows underabundances in the heavier elements. The average metallicity calculated from the elements Si to Ba is $M=-0.40\pm0.14$. If we include in the metallicity calculations all those species (C to Ba) for which we derived abundances, then we get a $Z$ value of 0.011. The He abundance was measured only from a single line (see Table \ref{abun}), but seems to be slightly higher than the solar value, i.e., by 0.04. In turn, this affects the $X$ value to be 0.72 instead of 0.74 as in the solar case (see Table \ref{models}). 

\begin{figure}[htb]
\centering
\includegraphics[width=0.45\textwidth,clip]{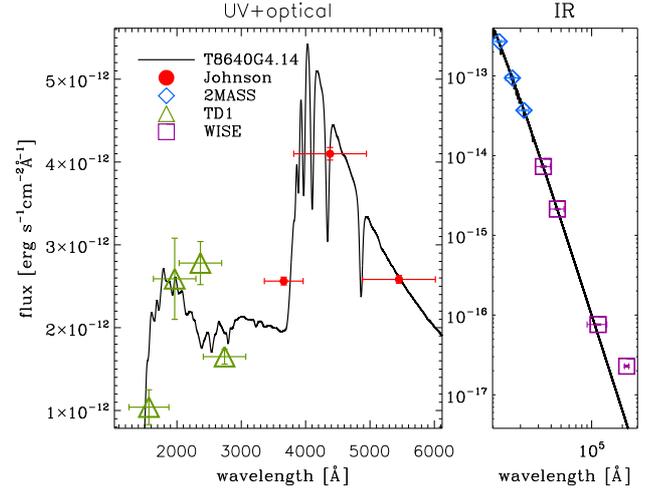}
\caption{Comparison between \llm\ theoretical fluxes (thick black line), calculated with the fundamental parameters derived for HD\,144277 (\Teff=8640 K, \logg=4.14), and observed Johnson (crosses), 2MASS (diamonds), TD1 (triangles), and WISE (squares) photometry converted to physical units.}
\label{sed}
\end{figure}

Fig.~\ref{sed} shows the fit of the synthetic fluxes, calculated with the fundamental parameters derived by us, to the observed TD1 (Thompson et al. \cite{TD1}), Johnson (Myers et al. \cite{Jonson}), 2MASS (Cutri \cite{2MASS}), and WISE (Cutri et al. \cite{WISE}) photometry, converted to physical units. Calibrations by Bessel et al. (\cite{bes98}), van~der~Bliek et al. (\cite{van96}), and Wright et al. (\cite{wri10}) were used for Johnson, 2MASS, and WISE photometry, respectively. The distance of 143$\pm$63~pc was taken from Kharchenko \& Roeser (\cite{kha09}) and the interstellar reddening $E(B-V)$=0.01 mag from Amores \& Lepine (\cite{AL}). The best fit was achieved for a stellar radius of $R$=1.56$\pm$0.69\,R$_\odot$; the relatively large error is mainly caused by the large uncertainty in distance.

For \Teff=8640~K and \logg\ of 4.14 an increase in gravity on the order of 0.15~dex results in a violation of the \ion{Fe}{i}/\ion{Fe}{ii} ionization balance exceeding the expected NLTE effects by a factor of two. We therefore take \logg=4.3 as a real upper limit. 
With the \Teff=8640\,K derived from the spectrum analysis and the radius of 1.56\Rsun, we estimate HD\,144277's luminosity to be log$L$/\Lsun=$1.09^{+0.38}_{-0.53}$ . 
The uncertainties in luminosity can only be used as formal errors that take into account the uncertainty of the distance fully. As HD\,144277 is certainly not a sub-dwarf, its radius must at least be larger than the solar radius. This would decrease the real lower error bar on the radius to 0.56\,R$_\odot$ and on the luminosity to 0.41.

A summary of the observationally determined values can be found in the third column of Table \ref{models}.

\section{Refinement of the asteroseismic model}

The asteroseismic model presented in Paper I was mainly derived based on information extracted from the observed frequency spectrum. With our new spectroscopic observations, we gained additional constraints on the fundamental parameters of the star. Consequently, the  asteroseismic model can now be tested and possibly refined. 

The new observations could confirm a lower metallicity compared to solar composition, which was predicted by the asteroseismic models in Paper I in order to explain the pulsational instability of the observed modes. The spectroscopically determined temperature, luminosity, and radius agree within one sigma to the values derived from asteroseismology, whereas the theoretical and spectroscopical $\log g$ values match only within $2\sigma$ (see Table \ref{models}). 
On the other hand, spectroscopy indicates that with $v \sin i = 62.0 \pm 2.0$\,\kms\ the rotational velocity of HD 144277 is higher than assumed in the asteroseismic models in Paper I, where an equatorial rotation velocity of 15\,\kms\ was adopted.
Since most fundamental parameters of the original asteroseismic model are in good agreement with the new observations, we used Model 2 from Paper I as a reference model (see Table \ref{models}) and increased its rotational velocity to study the influence on the predicted frequencies and mode instability. 

To estimate the effects of rotation on the frequencies we used a second-order perturbation theory, which is sufficient only for reasonably slow rotators. 
Increasing the rotational velocity at the equator from 15\,\kms\ to 60\,\kms\, implies that the effects of near-degeneracies gain importance. As shown in, e.g., Goupil et al. (\cite{goupil}) modes can couple if 
(i) they are close in frequency, (ii) their spherical degree differs by 2, and (iii) their azimuthal orders are equal. We therefore also considered the effects of rotational mode coupling according to the formalism presented in Soufi, Goupil \& Dziembowski (\cite{soufi}) and Daszy\'nska-Daszkiewicz et al. (\cite{das02}).
In our study we considered the rotational coupling of up to three modes and conducted tests thereby adopting various values for the rotational velocity. The adopted evolutionary and pulsational codes were the same as those used in Paper I. We again used OPAL opacities (Iglesias \& Rogers, \cite{iglesias}) and the Asplund et al. (\cite{asplund09}) proportions in the heavy element abundances to be consistent with Paper I.

\begin{figure*}[htb]
\centering
\includegraphics[width=0.85\textwidth,clip]{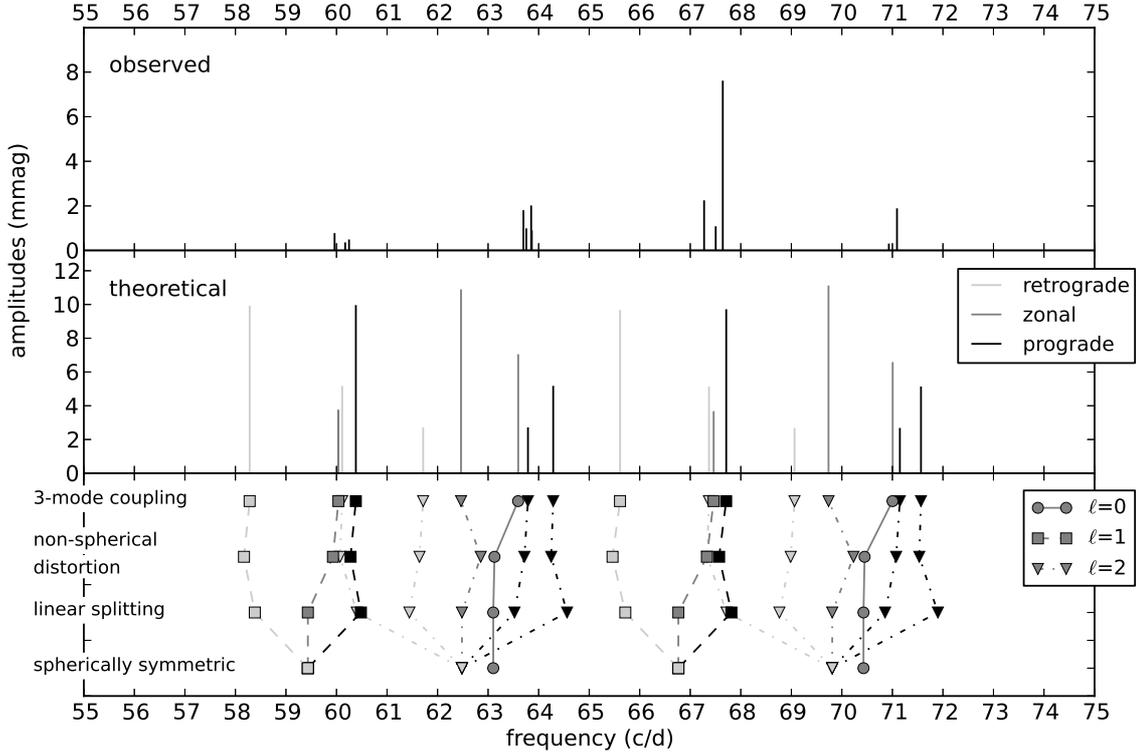}
\caption{{\it Top panel:} observed modes. {\it Middle panel:} theoretical modes with $\ell \leq 2$: retrograde (light grey), zonal (dark grey), and prograde (black) modes. 
{\it Lower panel:} Stepwise addition of various rotational effects on the mode frequencies in near-degenerate mode perturbation theory (spherical symmetry, linear splitting, non-spherical distortion, and three-mode coupling). Only $\ell$=0,1,2 are shown for clarity as filled circles, squares, and triangles. 
}
\label{splitting}
\end{figure*}

\begin{table}
  \caption{Parameters of asteroseismic models of HD\,144277 and comparison with the results from spectroscopy.}
  \label{models}
  \centering
  \begin{tabular}{llll}
    \hline
    \hline
    &Model 2 & Model 3 & Spectroscopy\\
    & (Paper I) & & \\
    \hline
     X & 0.70  & 0.70 & 0.72  \\
     Z & 0.010 & 0.011 & 0.011 \\
     M/\Msun & 1.66 & 1.70 & 1.72 \\
     log \Teff & 3.9407 & 3.9404 & 3.94 $\pm$ 0.01  \\
     log L/L$_\odot$ & 1.04 & 1.05 & $1.09^{+0.38}_{-0.53}$\\
     R/R$_\odot$ & 1.456 & 1.469 & 1.56 $\pm$ 0.69 \\
     log g & 4.332 & 4.325 & 4.14 $\pm$ 0.15\\
     V$_{\rm rot}$ ${[\mathrm{km\,s^{-1}}]}$ & 15.0 & 79.7 & (\vsini:) 62.0 $\pm$ 2.0 \\
    \hline
    \hline
  \end{tabular}
\end{table}

\begin{figure}[htb]
\centering
\includegraphics[width=0.45\textwidth,clip]{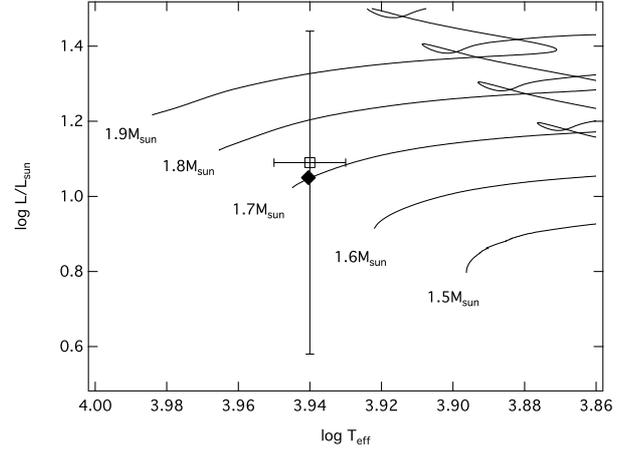}
\caption{HD 144277 in the HR-diagram: post-main sequence evolutionary tracks (solid lines) for 1.5 to 1.9\Msun\ in steps of 0.1\Msun\ computed for the same chemical composition (X, Z) as indicated by spectroscopy and by asteroseismology, refined asteroseismic model (solid diamond), and observational position (open square) of HD\,144277.  }
\label{hrd}
\end{figure}

We find that the best agreement between observational and theoretical frequencies is found if we assume a rotational velocity at the equator between 60 and 80\,\kms. For higher rotation there are no clear patterns in the theoretical frequency spectra that can be matched to the very regular pattern of the observed frequencies. This finding implies a nearly equator-on view. As can be seen in Fig.~\ref{splitting}, 
the observed groups of frequencies at 60 and 67\,\cd\ correspond to prograde and zonal dipole modes, while the other two groups are related to the radial and quadrupole modes.

\begin{figure}[htb]
\centering
\includegraphics[width=0.45\textwidth,clip]{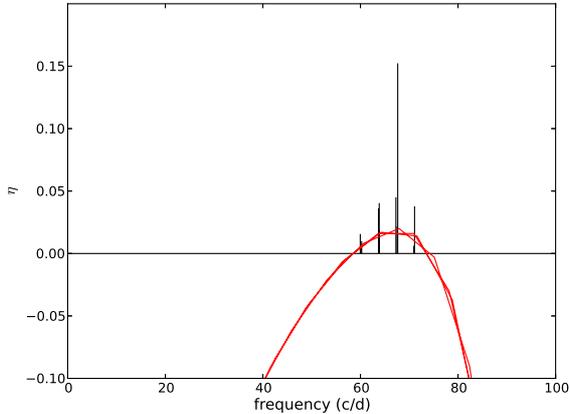}
\caption{Instability parameter, $\eta=\int_0^R(dW/dr)dr / \int_0^R|dW/dr|$, vs. mode frequency. Positive $\eta$ corresponds to driven modes. Observed modes are shown with vertical lines. The two (red) lines show the value of the instability parameter for $\ell$ = 1 and 2.}
\label{eta}
\end{figure}

Table~\ref{models} presents a comparison between the fundamental parameters of Model 2 from Paper I and the new model presented in this paper (denoted as Model 3). The position of the model in the Hertzsprung-Russell (HR)-diagram is shown in Fig.~\ref{hrd}. For the refined asteroseismic model a small increase in the metallicity was adopted to improve the agreement between the instability of theoretical modes and observed unstable modes (see Fig.~\ref{eta}). However, with Z=0.011 the model remains clearly metal underabundant compared to the solar value of Z=0.0134 (Asplund et al. \cite{asplund09}).

For stars with fundamental parameters similar to HD\,144277, the theory does not expect significant convective core overshooting. Indeed no overshooting is needed to derive the asteroseismic model presented in this paper; however, HD 144277 is too young for its observed frequencies and fundamental parameters to impose any real constraints on core overshooting to the model.

\section{Summary \& Conclusions}

Our first asteroseismic model for HD\,144277 was based solely on the observed pulsation frequency spectrum (see Paper I) since no high-resolution spectroscopy was available. This first model predicted a slightly less than solar metallicity and assumed a rotational velocity at the equator of only 15\,\kms. 

Using the high-resolution, high S/N HARPS spectra we confirmed the slightly less than solar metallicity needed to excite the observed \dsct\ frequencies in the range 59.9 to 71.1\,\cd\ (see Paper I). The fundamental parameters \Teff,  \logg, log L/\Lsun\ and R/\Rsun\ determined by spectroscopy agree within one sigma with the values found by asteroseismology. 

The major difference between the assumptions used in our first asteroseismic models and the spectroscopically derived values for HD\,144277 is the rotational velocity. While we assumed a rotational velocity at the equator of 15\,\kms\ in our first analysis, the projected rotational velocity, \vsini, is significantly higher at 62.0 $\pm$ 2.0\,\kms. In the present study we therefore included effects of near-degeneracies and rotational mode coupling of up to three modes. 

As the best agreement between the observed and theoretical pulsation frequencies is found for a rotational velocity at the equator of about 80\,\kms, the star has to be seen equator-on.

A near equator-on view implies that zonal modes have lower visibility than prograde and retrograde modes for geometrical reasons (see, e.g., Fig.~6 in Breger \& Lenz \cite{breger}). Our asteroseismic model indicates that the observed frequencies are prograde modes. Consequently, the following question arises: why are the retrograde dipole modes not observed, despite their having a similar geometrical visibility to the prograde modes?
A striking feature that may be the key to the physical mechanism of mode selection in the special case of HD\,144277 is that we particularly see those modes that are very close in frequency and affected by rotational mode coupling. These modes are in the center panel of Fig.~\ref{splitting} so that a better comparison can be made with
the theoretical mode distribution shown in the panel below and observations on top. 
This good agreement indicates that HD\,144277 may give us a valuable hint as to how the mode selection works in young $\delta$~Scuti stars.

\begin{acknowledgements}
The authors would like to thank our referee, Torsten B\"ohm, for his valuable comments that helped to improve the paper significantly.
We thank L. Mashonkina for providing us with the departure coefficients for the Mg NLTE analysis. 

The research leading to these results has received funding from the European Research Council under the European Community's Seventh Framework Programme (FP7/2007 -- 2013)/ERC grant agreement No 227224 (PROSPERITY). This research is (partially) funded by the Research Council of the KU Leuven under grant agreement GOA/2013/012. The research leading to these results was also based on funding from the Fund for Scientific Research of Flanders (FWO), Belgium, under grant agreement G.0B69.13.
KZ and MB acknowledge support from the Austrian {\it Fonds zur F\"orderung der wissenschaftlichen Forschung} (project P21830-N16). TR acknowledges partial financial support from Basic Research Program of the Russian Academy of Sciences ``Origin and Evolution of Stars and Galaxies''.
AAP acknowledges partial financial support from the Polish NCN grant No. 2011/01/B/ST9/05448. 
The INAF-OA Brera team acknowledges financial support from the PRIN-INAF 2010 {\it Asteroseismology: looking inside the stars with space- and ground-based observations.} MR also acknowledges financial support from the FP7 project {\it SPACEINN:
Exploitation of Space Data for Innovative Helio- and Asteroseismology}.
MH acknowledges support from the Austrian {\it Fonds zur F\"orderung der wissenschaftlichen Forschung} (project P22691-N16).
\end{acknowledgements}

\end{document}